\RequirePackage{fix-cm}
\documentclass[twocolumn]{svjour3}          
\smartqed  
\usepackage{graphicx}
\usepackage{hyperref}
\begin{document}

\title{Teaching Computational Neuroscience
}
\subtitle{Reviews:\\ \small{Britt Anderson: Computational Neuroscience and Cognitive Modelling
A Student's Introduction to Methods and Procedures. Sage, 2014\\
 Hanspeter Mallot:Computational Neuroscience: A First Course, Springer, 2013\\
Thomas Trappenberg: Fundamentals of Computational Neuroscience. Oxford Univ. Press, 2010
}}


\author{P\'eter \'Erdi        
}


\institute{P. \'Erdi \at 
Center for Complex Systems Studies\\
Kalamazoo College\\
Kalamazoo, MI 49006\\
1200 Academy Street\\ and \\
Institute for Particle and Nuclear Physics \\ Wigner Research Centre for Physics, Hungarian Academy of Sciences\\
              \email{perdi@kzoo.edu}   }       

\date{Received: date / Accepted: date}

\maketitle

\begin{abstract}
The problems and beauty of teaching computational neuroscience are discussed by reviewing three new textbooks.

\end{abstract}

\section{Teaching Computational Neuroscience: Diverse Perspectives}
\label{intro}
Computational Neuroscience is a discipline that has developed rapidly in the last twenty-five years. Roughly speaking it has two different meanings. First, how to use computational (more precisely theoretical and mathematical) methods to understand neural phenomena occurring at different hierarchical levels of neural organization. Second, how the brain computes (if at all). \\

While of course \emph{computational neuroscience} has its predecessors (from ''mathematical biophysics'', via ''cybernetics'' to ''theoretical neurobiology''), most likely the first book with the very title of ''Computational Neuroscience" was edited by Eric Schwartz \cite{schwartz90}. The chapters were written by celebrated authors and grouped into sections reflecting the hierarchical organization of the nervous system: Overviews, The Synaptic Level, The Network Level, Neural Maps, Systems. The flagship conference of the emerging discipline was organized by Jim Bower \cite{bower13} starting in 1992. As the Organization of the Computational Neuroscience website  (\url{www.cnsorg.org/}) describes: 
''\ldots Computational neuroscience combines mathematical analyses and computer simulations with experimental neuroscience, to develop a principled understanding of the workings of nervous systems and apply it in a wide range of technologies.''. \\

With a somewhat different perspective (shifting the emphasis on models from structural to functional aspects) Cosyne (Computational and Systems Neuroscience) conferences (\url{http://www.cosyne.org/}) have been organized since 2003 by Anthony Zador, Alex Pouget, and Zachary Mainen. As the conference website specifies: ''Cosyne topics include (but are not limited to): neural coding, natural scene statistics, dendritic computation, neural basis of persistent activity, nonlinear receptive field mapping, representations of time and sequence, reward systems, decision-making, synaptic plasticity, map formation and plasticity, population coding, attention, and computation with spiking networks. Participants include pure experimentalists, pure theorists, and everything in between." \\

There is a reemerging field of \textbf{mathematical neuroscience} with its own journal and conference (1st International Conference on Mathematical Neuroscience (ICMNS) in June 2015). I think this community will emphasize the necessity of using advanced mathematical methods, mostly for modeling less for data analysis purposes (but I might be wrong). The emphasis is on the the strict application of dynamic systems theory by using both deterministic and stochastic models.

Computational neuroscience  is a prototype of interdisciplinary science, and scientists following different traditions would agree that education is a key component in training   new generations of future experts, and helping experimentalists (biologists, clinical scientists, psychologists etc.) to understand what computational model can (and cannot) do.

In this paper three recent textbooks are reviewed. I am going to argue that while there is no reason to believe that it is possible or even necessary to write a single textbook, there are some overarching concepts and methods of computational neuroscience.  However, computational neuroscience is taught (i) to students, not only with different backgrounds, but also with different goals; (ii) by teachers with diverse backgrounds; (iii) by reflecting upon different intellectual forerunner disciplines; (iv) occasionally by emphasizing the importance of some specific levels of the neural hierarchy and (v) by emphasizing the importance of different mathematical methods.

\section{The Motivational Approach}
\label{sec:1}
\textbf{Computational Neuroscience and Cognitive Modeling} \cite{anderson14} is written by Britt Anderson, MD, a  psychologist at University of Waterloo. His motivation is clearly stated: ''I find many psychological undergraduates are interested in computational modelling, but that there is not a lot of literature or texts to help them get started. To try and improve that, I took the notes from a seminar I have taught over the last few years, and wrote a textbook.\dots ''. \\

Actually among the four parts of the book only the first two (1: Modeling Neurons, 2: Neural Networks) belong to computational neuroscience  in narrow sense, the other two parts (3. Probability and Psychological Models, 4.  Cognitive Modeling as Logic and Rules) should instead be taught in Cognitive Science classes.

\subsection{Modeling Neurons}
\label{ssec:m}

\emph{What is a Differential Equation?}\\	

This chapter is clearly written for those whom this is the first encounter with differential equations. It explains the difference between analytical and numerical solutions, by solving the algebraic equation $x^2=4$, and shows that an easy
computational technique of numerical (as opposed to analytic) calculations is a  spreadsheet program. Of course, students should understand what is a \textbf{D} in the \textbf{DE}, so derivatives as instantaneous slopes and the elementary jargon are discussed. \\

\noindent
\emph{Numerical Application of a Differential Equation}\\

There is a very simple, non-technical introduction to the idealized behavior of spring, i.e harmonic oscillator (Anderson mentions the joke with the physicists's ''spherical chicken'' approach). The example makes it clear that before doing anything students should understand the state variables of any system under study. In this simple mechanical example they are the \emph{velocity} and \emph{acceleration}. The most common  introduction to the methodology of solving differential equations numerically is Euler's method calculating the approximation of the derivative by using small, but finite changes in the variable. \\

\noindent
\emph{From Action Potential to Programming Neurons: Integrate and Fire}	\\

This chapter briefly introduces the concept of action potential; and the integrate and fire model.The explanation of the underlying physics is restricted to the definition of Ohm's law and Kirchoff's law. It uses a Java applet for solving the Hodgkin-Huxley equations in the condensed form Eq. \ref{eq:HHsimp}
\begin{equation}
\tau\frac{dV(t)}{dt}=RI(t)-V(t)
\label{eq:HHsimp},
\end{equation}
and explains the meaning of $dV/dt$, $\tau$, $I(t)$. The solution of the equation tells the evolution of the voltage, or membrane potential over time. Again, the author suggests to write a spreadsheet program to solve integrate and fire models. \\

\noindent
\emph{Hodgkin and Huxley: The Men and Their Model} \\

Part I about neurons is concluded by a chapter about the pioneers, and the emphasis is to explain \emph{conductance}-based models. A single current is substituted by the sum of individual currents flowing on specific ion channels, giving some hint to the interpretation of the gating variables. Concerning the educational level, there is an explanation in a nice box with the title ''What Are the Parentheses for?'', where the reader should understand that $V(t)$ is a number, actually the value of the function $V$ at the time-point $t$. \\

\noindent
\emph{Programming} \\

In addition to spreadsheet programming, Anderson suggests learning Python. While the book certainly cannot take the place of a programming class, it mentions some  elementary concepts of programming, such as  declaring variables, controlling structures  with loops, and conditions.

\subsection{Neural Networks}
\label{ssec:nm}

\noindent
\emph{Neural Network Mathematics: Vectors and Matrices} \\

How should we teach the elements of linear algebra for students without technical backgrounds? The first question, of course, Anderson addresses: ''What is a Vector?''. In addition to the statement that a vector is a list of numbers, there is also some hint about the geometric way of thinking on vectors. A number of basic concepts are mentioned, such as row and column vectors, with a smooth transition to the introduction of the concept of matrices and elementary vector and matrix operations, including inner products, dot product and matrix multiplication. \\

\noindent
\emph{An Introduction to Neural Networks} \\

The first question to be discussed is: ''What neural networks are?''. It is never easy to clarify the subtle connection between \textbf{biological}, real neural networks, and abstract \emph{artificial neural networks}. The author makes some references to the early history, starting with mentioning McCulloch and Pitts, John von Neumann and his cellular automata, and more generally the emergence of global structures from local interactions. It might be a good exercise for psychology majors to play a little with cellular automata. \\

One of the best parts of the book is the simple explanation of Perceptron. While it is true that one might decide not to teach any artificial network in a computational neuroscience class, its historical and conceptual importance justifies its place. The geometrical interpretation of the Perceptron learning rule helps students to understand how linear algebra works. The chapter concludes with the description of the delta rule, and with arguments about the necessity of having other neural networks, such as multilayer Perceptron and support vector machines. \\

\noindent
\emph{Auto-associative Memory and the Hopfield Net} \\

The Perceptron and its extensions were designed to solve classification problems. Looking back to its historical development, is it reasonable to ask about the  similarities and differences between Perceptron and Hopfield nets? Hopfield transferred a mathematical technique from physics (no more details were given). Essentially  some remarkable features are mentioned, such as error correction and the ability to complete patterns. Here the book became a little more difficult, students should understand some elementary concepts of dynamical systems, such as the algorithm for calculating  the outputs of a Hopfield network. Students should make exercises to understand asynchronous updating algorithms, and to convince themselves about the algorithm's convergence. \\
 
\noindent
\emph{Programming} \\

Octave programming language is suggested for learning how to write functions, and to implement and use delta functions.

\subsection{Cognitive Modeling}

Part III is about ''Probability and Psychological Models'', and Part IV is ''Cognitive Modeling as Logic and Rules''. While I see the importance of these concepts, in my own teaching experience it belongs to a Cognitive science class, and will not be addressed in this review.

\section{The Systems Neuroscience Approach}
\label{sec:sn}

Hanspeter Mallot' \textbf{Computational Neuroscience} \cite{mallot13} grew out of his notes on lectures delivered to his graduate students in neuroscience. The book provides an essential introduction to computational neuroscience at membrane, cellular, network and systems levels.

\subsection{Excitable Membranes and Neural Conduction}
\label{ss-m1}

The books starts with a description of membrane potentials, by mentioning but not writing  the Nernst equation. It continues with the Hodgkin-Huxley theory. Based on the  example of conductance change it is explained what is a differential equation and its solution. By explaining the physiology and kinetic models of potassium and sodium channels of the original (spatially homogeneous) Hodgkin-Huxley model, a 4D ODE is specified, and a hint is made that such kinds of equations can only be solved numerically. The two-dimensional approximation of the model (the FitzHugh-Nagumo model), or at least its qualitative properties can be studied analytically. The book does not go into detail about how changes in the input parameter lead to transitions from equilibrium to oscillatory response.

It is always difficult to decide how to teach concepts such as the spatio-temporal propagation of action potentials. The explanation requires some knowledge of basic concepts of partial differential equations. There is a section on passive conduction to give an intuitive introduction to the cable equation, and to combine this equation with the kinetic equations to channel models

\subsection{Receptive Fields and the Specificity of Neuronal Firing}
\label{ss-m2}

How neurons integrate activities in space and time is an important topic for understanding the functions of specific neurons. The superposition principle states that stimuli coming from different space points can be added. The rigorous understanding of the concept of the \emph{receptive field function} $\Phi(x,y)$ requires advanced math, such as the theory of linear operators, which belongs to field of functional analysis. The phenomenon of lateral inhibition, i.e. the reduction of the activity of neighboring cells already demonstrated  by Mach, and it is related to the notion of \emph{convolution}. Its understanding needs some insights about integral equations. Tuning curves describe the response of a neuron for different stimulus parameters. \\

Simple receptive fields are isotropic  or show rotational symmetry, and they are well modeled by the Gaussian
function. To describe orientation-dependent effects the celebrated \emph{Gabor function} is extensively used. The \emph{Gabor function} is generated by multiplying sinusoidal and Gaussian functions. \\

The relationship between the set of all stimulus and  the set of possible excitatory responses in the general case is described by nonlinear operators. Static nonlinearities in receptive fields  (such as threshold, saturation and  compression) are illustrated. There is some explanation of the concept of Volterra kernels, for purely temporal systems. \\

Motion detection is identified to calculate velocity and direction. The computational problem was suggested to be solved by designing coincidence  and/or correlation detectors. From and educational perspective it requires the knowledge of some elementary probability theory. The author jumps suddenly to introduce auto- and cross-correlation functions, and hopefully most students have the background to understand them. It might have been a reasonable question to ask what type of computations can be implemented by single cells, and what should remain for neural networks.

\subsection{Fourier Analysis for Neuroscientists}
\label{ss-m3} 

I read this chapter with great interest . It starts with examples of light spectra, acoustics, vision and magnetic resonance tomography. Terms  such as ''spectrum'' and ''Fourier decomposition'' appear. Mathematical concepts from complex numbers to Gaussian convolution kernels have been reviewed. Eigenfunctions and eigenvalues are introduced as characteristics of the operator connecting input images and activity distributions. The eigenfunctions of convolution is explained in detail, using both real and complex number notations.

The basic theory of Fourier decomposition is based on the fact that most functions can be expressed with the linear superposition of sine and cosine functions. Via the convolution theorem the construction leads to  algorithms for finding the coefficients of the Fourier series.  Generalizations for non-periodic functions and for higher dimensions were also given.


\subsection{Artificial Neural Networks}
\label{ss-m4}

I decided not to review ANN in detail. From an educational point of view the mathematics explained here is the  dot product and matrix operations. Two fields of applications, namely classification and associative memory are discussed.

\subsection{Coding and Representation}
\label{ss-m5}

This chapter contains two parts, population code and retinotopic mapping.\\

There are different possibilities of coding with neural activity. Intensity code means that the activity of a neuron is a monotonous function of a coded parameter. Channel coding and  population coding are used more frequently. While population code needs more neurons to encode some parameter value of a stimulus, it still looks superior to others, e.g. it leads to better resolution.

The section about information content of population codes is used to explain the basic notions of information theory. The celebrated concept of the center of gravity estimator for reading a population code is described.\\

The last section about retinotopic mapping is written based on the research the author did himself with Werner von Seelen (who actually wrote the Foreword of the book) in the 1980s. In terms of mathematics the problems  is describing the coordinate transformation from retina to visual cortex. Conformal maps and log-polar mapping are the appropriate tools to describe the geometry of retinotopy.

\section{The Computational Approach}
\label{sec:cs}

Thomas Trappenberg's \textbf{Fundamentals of Computational Neuroscience} \cite{ttrap10} is a comprehensive textbook for advanced graduate students, It tries to teach the minimally necessary neurobiology that a computational science student should know. In terms of using mathematics the author takes a pragmatic approach, mentioning mathematical concepts when explicitly used. Some elementary textbook notions of linear algebra, calculus, numerical solutions, probability and information theory and some introduction to  MATLAB were put into the Appendix. I found the book very carefully and clearly written and its spirit is close to that I am using in classes. \\

The book has three main parts (Basic Neurons, Basic Networks, System-Level Models) following the spirit of bottom up modelling, preceded by a nice introduction to computational neuroscience. Of course, it is somewhat a matter of personal taste how to write briefly about such questions as ''What is computational neuroscience?'', ''What is a model?'', ''Is there a brain theory?'', and ''A computational theory of the brain''.

\subsection{Basic Neurons}

This part starts with a chapter on neurons and conductance-based models, with a little biology about cellular morphology, synapses, ion channels and membrane potential. After writing about synaptic mechanisms, it explains the problem of modelling synaptic response, maybe a little too early. This section seems to be a little dense, and between the explanation of Kirchhoff's law and Ohm's law there is a short introduction to differential equations. The next sub-chapter explains the generation of action potential and the Hodgkin-Huxley equations followed by some elements of numerical integration, and its implementation with MATLAB. Cable theory uses partial differential equations (actually it is a parabolic, and not a hyperbolic equation, as it is written). The spatial discretization of neurons leads to compartmental models, and the reader get some reference to the two most frequently used neuron simulators (GENESIS, NEURON). \\

The next chapter comprises two reasonable topics: why and how to simplify neurons, and how to treat the dynamics of neural populations. Trappenberg writes correctly: one trivial reason is to make possible calculations with large number of neurons, but conceptually more importantly to get an insight into the skeleton mechanism of the generation of emergent network properties. It is somewhat arbitrary, what is a good order to explain different models. Here first we see  leaky integrate-and-fire neuron, then
 spike response model, Izikievich neuron and McCulloch-Pitts neuron. (I like to teach following the historical development, teaching McCulloch-Pitts model first.) One challenge is to explain the the response of neurons for different (constant, and time-dependent) inputs. While the integrate-and-fire models neglect the details of subthreshold dynamics, the Izikievich neuron is a nice compromise between biological plausibility and computational efficiency. All the previous models are deterministic, and cannot describe the inherent random character of neural firing. Somehow stochasticity should be put to the models. The book actually does not mention the methods grew up from the classical paper of Gerstein and Mandelbrot \cite{gerstein64} which consider the membrane potential change as a random walk. Spike-time variability can be explained with the introduction of stochastic threshold or reset. The noisy integrator puts the randomness to the input. \\

One of the main challenge in theoretical neurobiology is to understand the nature of neural code: what is the relationship between firing patterns and behavior. There is a somewhat different question about the relationship between sensory stimuli and neural activity patterns. Rate code (going back to Lord Adrian's finding), correlation codes and coincidence detectors are explained, though maybe a little more about temporal coding is missing. \\

While  we are still discussing single neurons  (more or less: there is a section about population dynamics   to be studied again in a later chapter). The state of the population of neurons can be characterized by some population activity vector. Actually the vague concept of \emph{cell assembly} (or \emph{local pool}, based on the terminology used in the book) suggests that cells in a certain pool have common features. I miss something more about mesoscopic neurodynamics. (We worked slowly twenty years ago on a model of large populations of neurons. What we saw is that a ''mean field'' approach is not enough, and the behavior at least as for an ''average cell'' - which is a hypothetical neuron receiving the average synaptic input calculated by a statistical approach - is also necessary.)  The chapter  finishes with two clearly written sections; one describes the different activation (or transfer) functions, and the other networks with non-classical synapse taking into account nonlinear interactions among neurons.\\

What might be more complicated than one neuron? Two neurons connected by a synapse. While the connectivity matrix of a network is able to store memory traces, learning is related to the mechanism of synaptic modifiability to be described by (generalized) Hebbian learning algorithms. As the book correctly reflects, Hebbian learning mechanisms first explained  psychological (and not neural level) phenomena, such as learning by associations, and conditioning. Going downwards on the ladder, then the physiological basis of plasticity is given. As it is well known, the identification of the cellular mechanism of synaptic plasticity (related to the long term potentiation and long term depression) was given a quarter century later after the proposal of  the Hebbian mechanism. Actually for me it was very useful to learn about the work of Alan Fine and Ryosuke Enoki about their plasticity experiments by combining paired-pulse facilitation and calcium imaging of single synapses. At even lower level, the biochemical pathways related to the role of NMDA receptors and calcium ions is discussed. \\

While Hebb  described his rule verbally,  there are now many variations of mathematical forms of the phenomenological learning rules. Again, as a matter of taste, Trappenberg starts with a newest rule (spike timing dependent plasticity); the book does not mention what it is now well known:  that there was an early discovery of the phenomenon by Levy and  Steward 1983 \cite{levy83}. I am comfortable with the way of presentation (but we did a different approach in the second edition of Michael Arbib's wonderful Handbook of Brain Theory \cite{arbib02}). Also, the last (very well written) section on synaptic scaling and weight distributions should have been written within the section on mathematical formulation. 

\subsection{Neural networks}

There is a very difficult question without a clear answer: What to teach about neural organization? Trappenberg discusses large-scale neuroanatomy and hierarchical organization, the layered structure and the modular organization of the cortex (according to my biased view, John Szent\'agothai (and also Valentino Braitenberg) might have been mentioned). \\

I am not totally sure whether random networks should be studied emphatically Actually, Anninos and his colleagues are mentioned, but more precisely they considered neural networks with features described as ''randomness in the small and structure in the large''.
The remaining part of the chapter about more physiological spiking networks based on Izikievich is plausible, and appropriate for demonstrative simulations. From an educational perspective it is not convincing that spectral analysis and fast Fourier transformation mentioned without telling anything about them.\\

The chapter about feed-forward networks belongs mostly to artificial neural networks, and according to the spirit of this review I don't discuss it in details. The single and multilayer perceptron and support vector machine are clearly presented for classroom teaching. \\

Cortical maps between two regions are used to represent features. One of the main principles of neural organizations states that maps often have a topographic character, i.e. neighboring areas represent neighboring features.The normal ontogenetic formation and the plastic response of these maps for certain lesions have a common  mechanism  modeled by self-organizing algorithms. The famous Willshaw - von der Malsburg model and the Kohonen algorithm are the paradigmatic examples of these algorithms. Maps get newer training input patterns, but they should preserve their basic properties. The general problem ''how can a system retain old memories but learn new ones'' is called as the stability-plasticity dilemma \cite{grossberg80}. \\

In certain applications it is plausible to use continuous space coordinates instead of discrete neurons, so the appropriate dynamic model to describe spatio-temporal activity propagation would be a partial differential equation. In the general case the system is non-autonomous, since the input is time-dependent. However, the activity dynamic at a certain point depends on other spatial coordinate  taken into account by an integral (kernel). Gaussian function and the Mexican-hat functions are  characteristic examples of centre-surround interaction kernels. The dynamics lead to different qualitative states. One of them is characterized by a stable active region even after the removal of external input, so memory can be stored. Mathematically the construction leads to continuous attractors. Again, from an educational perspective it is disadvantageous that we have not yet read about the simpler concept of point attractors. Inferior temporal cortex, prefrontal cortex implementing working memory and hippocampus' involvment in space representation are illustrative examples of the computational concept. \\

The main goal of the chapter (Recurrent associative networks and episodic memory) is to show how recurrent neural networks might be the anatomical substrate of  associative (specifically auto-associative) memory. Auto-associative memory basically does pattern completion. The reentrant connections among the pyramidal cells of the hippocampal  CA3 region  implement a network capable of storing and retrieving, (at least static) input patterns. While there were forerunners and followers, John Hopfield \cite{hopfield82,hopfield84} popularized an abstract network implementing point attractors mostly for the physicist's community (and it cannot be denied that Google Scholar gives altogether twenty-thousand  citations for the two papers.).\\

Section 8.2 nicely describes the standard math coming from the physicists theory of phase transitions of the spin glasses. A remarkable feature of the attractor neural networks is their noise tolerance of the stored memories. While the original work (and the underlying mathematical theorems) are based on rather restrictive assumptions, simulation results generally permit the extension of the theory. In this book some important further studies, such as the effects of dilution (i.e the deviation from the fully connected networks), and for non-random, but correlated, input patterns are discussed. There is one neglected problem with attractor neural networks, that Trappenberg also does not mention: real neural systems processes time-dependent inputs, so the differential equations set to describe their dynamic behavior are non-autonomous. Non-autonomous systems don't have attractors in the general case.\\
 
I find it somewhat arbitrary both the ''where'' and the ''how'' of  discussions of the dynamic systems and chaotic attractors. I might be biased, but I think a separate chapter on the dynamic approach would be useful. Between the point attractors and chaotic regimes the very important neural oscillators should have been studied. The Cohen - Grossberg theroem is extensively described. I would have liked to see  Hirsch's celebrated paper \cite{hirsch89}, which gives a more generalized but still operative study of the problem. Also, both the structural basis and the functional significance of the chaotic neural network (see e.g \cite{tsuda92} might have been discussed.

\subsection{System-Level Models}

Section nine (Modular networks, motor control, and reinforcement learnig) and ten (The cognitive brain) are about system level models, and I find them somehow the less convincing part of the book. I acknowledge the important aspects of the discussed topics. Modular mapping networks, coupled attractor networks, sequence learning, complementary memory systems, motor learning and control, reinforcement learning are all important topics. However, I somehow  missed  the coherent story line. Mixture of experts is clearly a topic in machine learning. "Sequence learning", gave a partial answer about what to do with (at least some specific) time-dependent inputs. It would have been possible to show either motor sequence learning, or to discuss the role of the hippocampus in sequence learning. \\

Obviously it is very difficult to write a short chapter about the cognitive brain. What should be the illustrative examples are a matter of taste. The first two examples Trappenberg chose are attentive vision and the interconnecting workspace hypothesis. I find very important and also well-written the section about the \textbf{anticipating brain}. A large school of computational neuroscience adopts the top down computational approach. The starting point is that the brain builds a model of the external world and generates hypothesis about the consequences of possible actions. The formal framework to formulate the predictive power is probabilistic. The subsection on probabilistic reasoning and Bayesian network briefly discusses a technique for analysis. The main corpus of the book is ended by the adaptive resonance theory, Any author has the right to decide how to finish a book.





\section{Computational Neuroscience: A Wonderful Subject to Teach and Learn}

The discipline of Computational Neuroscience discusses mathematical
models, computational algorithms, and simulation methods that
contribute to our understanding of neural mechanisms. Students can learn
the basic concepts and methods of computational neuroscience
research, and absorb some knowledge of neurobiological concepts and
mathematical techniques. \\

What are the goals of teaching computational neuroscience? The first goal is to teach WHY mathematical and computational methods are important in understanding the structure, function and dynamics
of neural organization. The second goal is to explain HOW neural
phenomena occurring at different hierarchical levels can be described
by proper mathematical models. \\

The books under review took some different perspectives, had different goals, and it is good to see that teachers have alternatives. and there are not the end. There are a couple of other books around, I simply selected books, which for one or another reasons were on my desk. Copies of \cite{dayan-abbott01} are on my shelves, both in Budapest and Kalamazoo, and this book is written by two very highly respected scientists. The book has three parts (Part I: Neural Encoding and Decoding; Part II: Neurons and Neural Circuits; Part III: Adaptation and Learning). When I used this book in an my undergrad class, I felt I should skip the first part, and was able to teach the large part of  Part II, and a few examples from the Adaptation and Learning Section. There are at least two more books written by equivalently highly respected colleagues I should have reviewed. \cite{ermentrout} and \cite{izhikevich}  which clearly represent the high level state-of-art of mathematical neuroscience. Another new great book \cite{gerstner14} is also left  for another time and maybe for another reviewer.

\begin{acknowledgements}
Thanks for my numerous  teaching assistants over the years. I had many conversation with them about the method of teaching of this discipline. I also thank to the Henry Luce Foundation to let me to serve as a Henry R Luce Professor. Thank you for Brian Dalluge (who is now in my Computational Neuroscience class) for copy editing the manuscript.
\end{acknowledgements}

\bibliographystyle{spmpsci}      
\bibliography{cnsrev}   

%
%

\end{document}